# Nanoimprinted topological laser in the visible

Qiang Zhang[1,#], Rui Duan[2,#,*], Yutian Ao[1], Lin Wang[3], Xuehong Zhou[1], Xuyong Yang[3], Xiao-Cong Yuan[4], Baile Zhang[1,5,*] and Handong Sun[2,*]

[1] Division of Physics and Applied Physics, School of Physical and Mathematical Sciences, Nanyang Technological University, Singapore 637371
[2] Institute of Applied Physics and Materials Engineering, University of Macau, Macao SAR 999078, China
[3] Key Laboratory of Advanced Display and System Applications of Ministry of Education, Shanghai University, 149 Yanchang Road, Shanghai, 200072, China
[4] Nanophotonics Research Center, Shenzhen Key Laboratory of Micro-Scale Optical Information Technology, Institute of Microscale Optoelectronics & State Key Laboratory of Radio Frequency Heterogeneous Integration, Shenzhen University, Shenzhen, China
[5] Centre for Disruptive Photonic Technologies, Nanyang Technological University, Singapore 637371

[#]These authors contribute equally

* e-mail: ruiduan@um.edu.mo, blzhang@ntu.edu.sg and hdsun@um.edu.mo

**ABSTRACT**:

Nanoimprint lithography (NIL) is a widely used high-throughput fabrication technique for photonic devices, yet its reliability is often compromised by the inevitable imperfections that arise during the demolding process. Topological photonics, which harnesses topologically nontrivial structures to support defect-robust photonic states, offers a promising solution to this challenge. Here, we demonstrate a topological laser that is one-step nanoimprinted upon colloidal perovskite nanocrystals. This laser features multiple higher-order topological corner states, with topological protection provided by the structure effectively mitigating imperfections caused by the nanoimprinting process. This property enables the reliable detection of these states, which is particularly challenging to achieve in the visible spectrum. Our work establishes topological photonics as a viable pathway to enhance the reliability of NIL-based manufacturing, providing a scalable and practical route for mass-producing topological lasers with low-index materials.

Nanoimprint lithography (NIL) is a low-cost, high throughput fabrication technique that employs a soft stamp or mold [1], which can be repeatedly used to transfer the same pattern onto multiple substrates. This technique is particularly well-suited for flat optics applications with on-chip integration. NIL offers two key advantages: (i) it is significantly more cost-effective than conventional fabrication methods such as electron beam lithography and focused ion beam milling, and (ii) it enables the mass production of large-scale devices. However, a major challenge in developing practical visible-wavelength nanodevices is the stringent fabrication tolerances required [2-9]. This limitation presents a significant hurdle for NIL in active devices operating in the visible range, where fabrication imperfections inevitably arise during the demolding process. This issue is further exacerbated when using solution-processed functional materials, such as organic dyes and colloidal semiconductor nanocrystals (NCs).

Topological photonics [10, 11] offers a promising solution to this issue. Topology is a mathematical concept that characterizes properties of objects that remain invariant during continuous deformation. Just as it has been used in condensed matter physics to describe topological phases of matter, such as topological insulators [12, 13], its use in photonics has led to the emerging field of topological photonics. This field explores topological photonic states that are inherently robust against certain classes of fabrication imperfections and disorders [14-17]. Among the various applications of topological photonics [18, 19], one of the most promising is the development of topological insulator lasers, or topological lasers, where coherent light emission is protected by topological effects [20-35]. However, to date, most topological lasers [20-23] [25-29] operate in the near-infrared (IR) regime with III-V group gain materials, requiring precise nanofabrication techniques that are not well-suited for large-scale production.

Here we integrate NIL with topological photonics to enable the development of cost-effective, large-scale topological photonic devices, including topological lasers. As a proof of concept, we demonstrate a room-temperature nanoimprinted topological laser operating at a wavelength of 523 nm. The gain medium consists of all-inorganic colloidal caesium lead halide ($CsPbBr_3$) perovskite NCs, chosen for their superior optical gain properties and solution processability [36]. The device design is based on a two-dimensional (2D) photonic lattice supporting higher-order topological corner states (HOTCS), as illustrated in Fig. 1a. Corner states have been extensively studied across various platforms [30-33] [37-64]. Considering only nearest-neighbor (NN) coupling, type-I corner states arise at the lattice vertex and are pinned to zero energy [30-33] [37-55]. Introducing next-nearest-neighbor (NNN) coupling gives rise to type-II corner states, which are localized at the second cells from the lattice vertex [56, 61, 62]. Increasing the NNN coupling strength further leads to type-III corner states, where wave functions localize at the third cells from the lattice vertex [63, 64]. While previous observations have primarily focused on type-I and type-II corner states, leaving type-III states largely unexplored-- especially in the optical regime [65, 66] - here, we demonstrate the modulation of type-III corner states with distinct symmetries via parity engineering. This is achieved by manipulating the parity of long-range interacting units. Notably, these topological corner states are demonstrated in lasing actions, exhibiting remarkable frequency stability even in the presence of significant fabrication imperfections.

Our work provides both theoretical and experimental insights into the photonic type-III HOTCS, along with a topological laser that harnesses this state. This highlights the clear advantages of integrating NIL technology with topological photonics, offering superior robustness and performance compared to conventional NIL-based approaches.

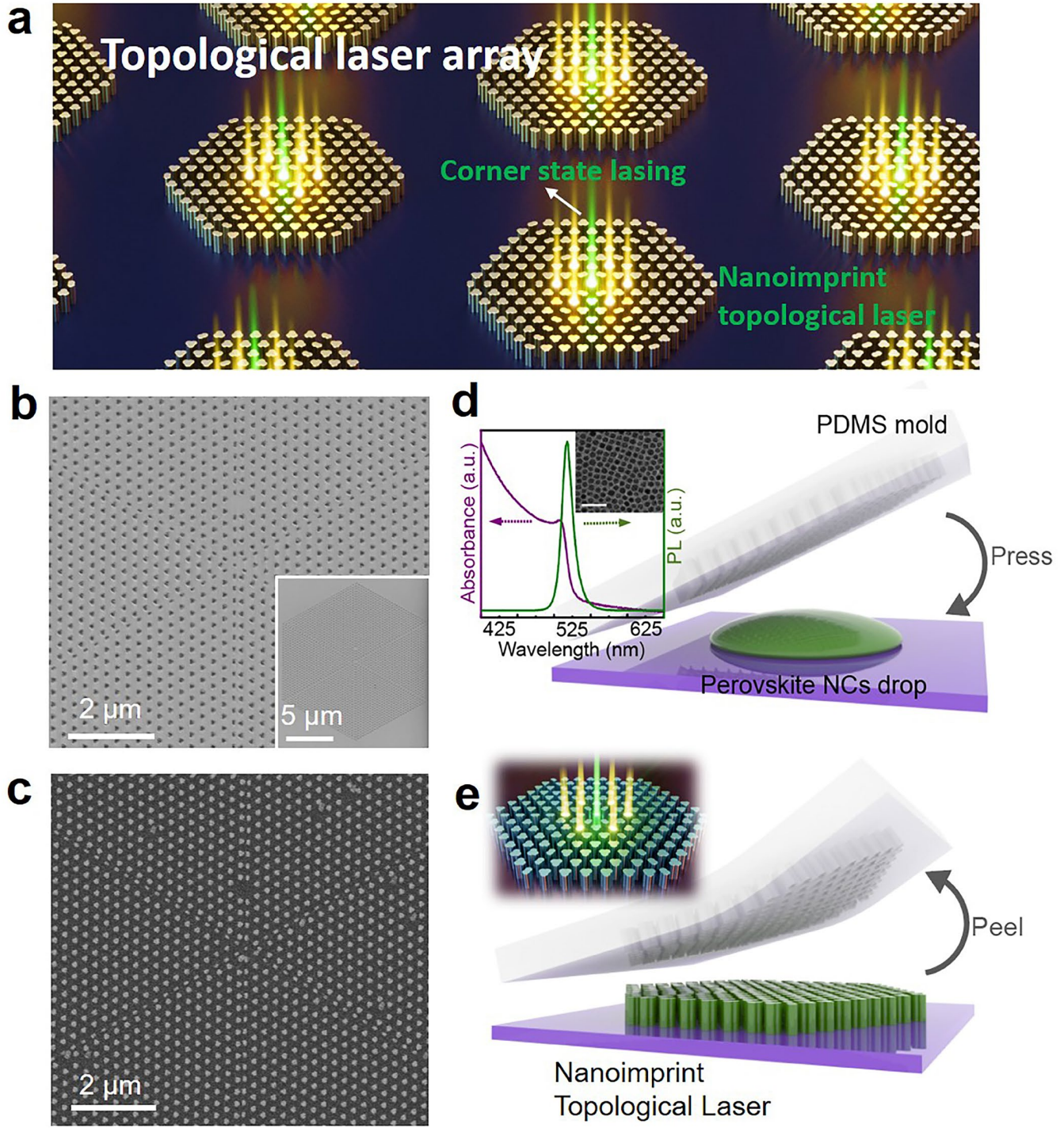

**FIG. 1. Nanoimprinted topological lasers. (a)** Illustration of topological lasers exhibiting multiple corner states. The topological cavities are fabricated on a distributed Bragg reflector (DBR) which is highly reflective in the green light range. **(b)** Scanning electron microscope (SEM) image of the original silicon master. Note the photoresist residual are left in the holes after pattern transfer to a working master. **(c)** The final nanoimprinted higher-order topological corner state laser. Fabrication imperfections exist due to the manual demolding process. **(d)** Schematic of the templated-assembly nanoimprint lithography (NIL) method. Imprinting: a soft PDMS mold is pressed onto a drop of colloidal $CsPbBr_3$ perovskite NCs on the DBR. The top-left inset shows the absorbance, PL spectrum and TEM image of the nanocrystals (NCs). **(e)** Demolding: after drying, the mold is carefully peeled off, leaving the perovskite nanopillars on the DBR. The soft PDMS mold is a negative replica of the photoresist working master. The original master consists of nanoholes - the inverse patterns of the nanopillars.

## Results

**Fabrication and design principle**

The NIL process is schematically illustrated in Fig. 1. The inset of Fig. 1d shows the photoluminescent (PL) spectrum of the NC solution, featuring a peak at 517 nm and a full width at half maximum (FWHM) of 16.5 nm. The photoluminescent quantum yield (PLQY) of the NCs was measured to be as high as 0.96, providing a gain foundation for the low threshold of the topological laser. Specifically, a drop of perovskite NCs solution, with hexane as the solvent, is added onto a highly reflective distributed Bragg reflector (DBR), as depicted in Fig. 1d. The DBR is designed to operate in the green light range, with a reflectivity exceeding 0.98, to enhance the quality factor of the topological cavity. After drop-casting, a PDMS mold is immediately pressed onto the solution. Once the solvent has completely evaporated, the mold is carefully peeled off, leaving nanostructured perovskite patterns on the DBR (Fig. 1e). The PDMS mold is a negative replica of a working master template made from photoresist, which in turn is a negative replica of the original silicon master stamp. A key advantage of both the PDMS mold and the working master is that they can be reused for many times. The original silicon master stamp consists of nanoholes (Fig. 1b), which are the inverse patterns of the nanopillars. Notably, residual photoresists from the preparation of the photoresist working master led to other defects in the negative replica of the original master. The fabricated sample is demonstrated in Fig. 1c. Although fabrication imperfections are evident in the sample during the manual process, the topological lasing action remains stable, as demonstrated later. This robustness against fabrication defects highlights the superiority of topological lasers.

The proposed structure consists of three pairs of alternating trivial and topological (non-trivial) Kagome nanopillar lattices (Fig. 2a). We align the three topological Kagome lattices to share a single apex nanopillar, as illustrated in Fig. 2a. The trivial and non-trivial phases are realized by shrinking and expanding the unperturbed lattice, respectively, as shown in the unit cells in Fig. 2a. This process adjusts the inter-cell coupling strength ($t_b$) and intra-cell coupling strength ($t_a$). These nanopillars support transverse magnetic modes. In an unperturbed Kagome lattice, the inter-cell and intra-cell coupling are identical, resulting in a Dirac-like degeneracy at the K (K') point in the band diagram (*SI Appendix*, section 2). When the balance between the two couplings is disrupted, a complete bandgap opens. The expanded lattice ($t_b > t_a$) and the shrunken lattice ($t_b < t_a$) share the same band structure, but their bulk polarizations, which characterize the topology of the lowest energy band, differ significantly [50, 56].

As previously demonstrated, a single non-trivial Kagome lattice embedded within a trivial lattice hosts HOTCS, pinned to zero energy at the fundamental resonance frequency of a single nanopillar [51, 52], i.e., the type-I corner state. Moreover, when next-nearest-neighboring (NNN) coupling is considered, a type-II corner state emerges, where the intensity becomes localized at the two dimers closest to the apex corner [56]. This occurs due to non-negligible long-range interactions, which are common in electromagnetic wave systems. In our design, the nanopillars of the expanded Kagome lattice overlap with those from neighboring cells, thereby increasing the NNN coupling strength to a level comparable to $t_a$. This configuration is expected to induce the formation of type-III corner states.

Based on the symmetry of the $E_z$ component profiles of the electric field relative to the angle bisector of each corner, we classify the different types of corner states into antisymmetric-type-II (A-type-II), symmetric-type-II (S-type-II), antisymmetric-type-III (A-type-III) and symmetric-type-III (S-type-III). To investigate the symmetries of the four type-II and type-III corner states, we developed an effective model beyond dimer couplings in the Kagome lattice. Previous studies have shown that the A-type-II corner state is well described by direct dimer-dimer coupling at the corners [56], and the S-type-II corner state is characterized using a dimer-trimer coupling model [62]. While stronger dimer-trimer coupling [63] or NNN coupling [64] can induce type-III corner states, a model specifically addressing their mode symmetry has been lacking. We show that the trimers can be treated as electric monopoles and dipoles when engaging in the dimer-trimer interactions. By considering these scenarios, we derive rectified Hamiltonians for all types of HOTCS (*SI Appendix*, section 1), which allow us to obtain more precise energy spectra and further characterize their symmetry features.

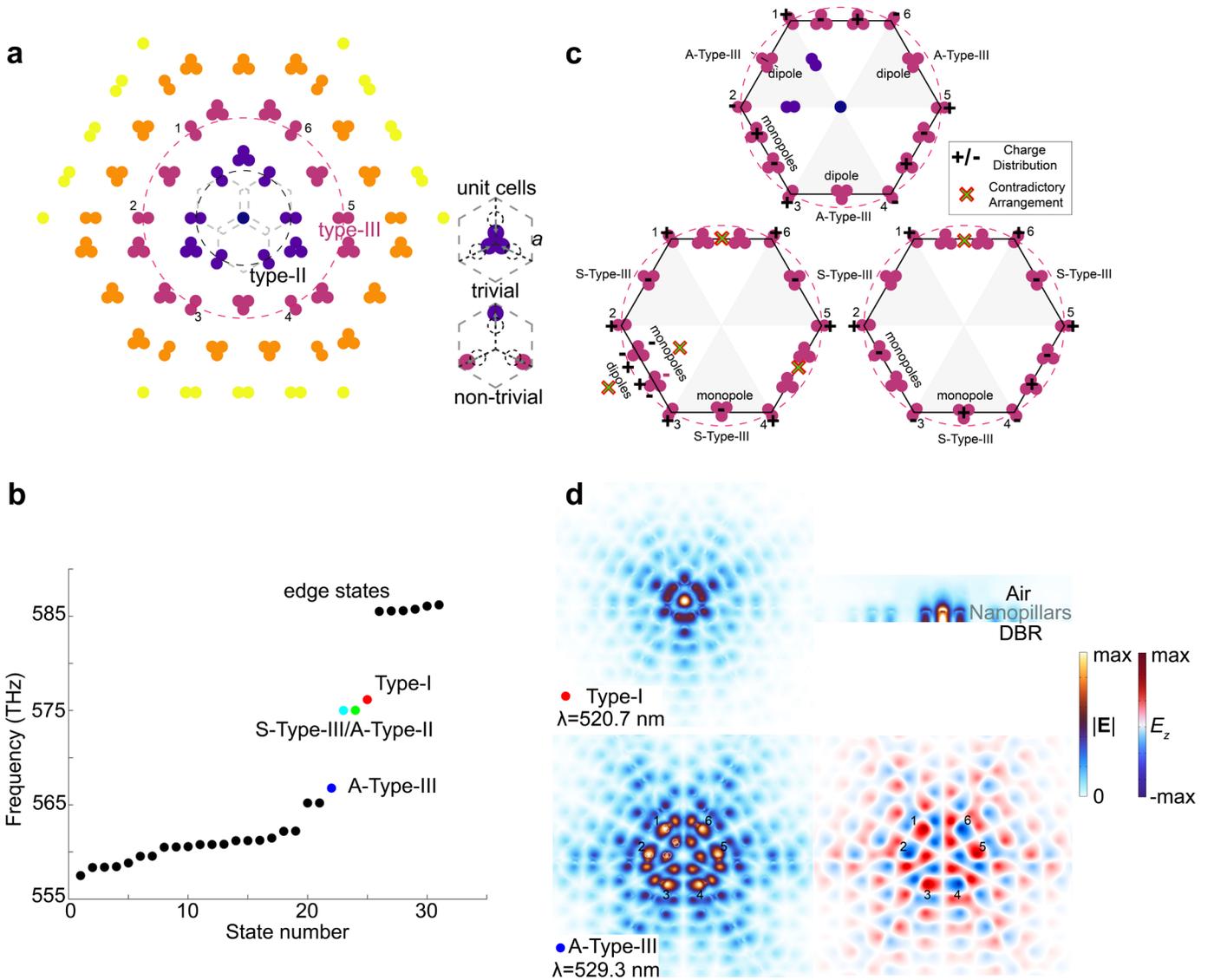

**FIG. 2. Topological nanocavity.** (a) Schematic diagram of the proposed topological cavity, composed of three trivial and three non-trivial Kagome perovskite nanopillar lattices. The trivial and non-trivial lattice are realized by shrinking and expanding the unperturbed nanopillar lattice (dashed circles in the insets). The three non-trivial lattices are translated toward the center with a small displacement such that their lattice vertexes can overlap to share one pillar (coincide). The dashed circles highlight the locations of type-II (dimers at the second cells from the vertex along the boundary) and type-III corner states (dimers at the third cells from the vertex along the boundary), respectively. Each dashed circle passes through six dimers, labeled from 1 to 6. (b) First-principle 3D numerical simulation of the band structure of the topological cavity following the design in (a). Different corner states are marked by different color dots. S-type-III and A-type-II modes are almost mixed due to the narrow bandgap. (c) Charge distribution and coupling mechanisms in a system with type-III corner state formed in six dimers arranged along a dashed circle trajectory. For the antisymmetric mode, dimers exhibit alternating positive and negative charges, with trimers acting as electric monopoles or dipoles to mediate coupling (top panel). In symmetric charge distribution scenarios (Scenario 1: bottom left; Scenario 2: bottom right), parity constraints prevent valid monopole or dipole couplings, as indicated by cross marks. These constraints enable the A-type-III corner state to remain spectrally isolated, while the S-type-III corner state tends to mix with type-II mode due to the parity incompatibility. (d) Field profiles of the corresponding type-I corner and type-III corner states.

We performed 3D first-principle finite element method (FEM) simulations, assuming a refractive index of 2.0 for the nanopillars. The nanopillars are set with a height of 170 nm, while the original silicon master stamp contained holes with a depth of 190 nm-- slightly deeper than the actual nanopillar height - to accommodate residual air bubbles that might remain trapped during the drying process. The simulated eigenmodes are presented in the band structure diagram shown in Fig. 2b, where hierarchical corner states are highlighted with dots of different colours. Notably, the bandgap is further compressed compared to the 2D case (*SI Appendix*, section 2). A type-III corner state is identified, which, despite recent observations in electrical circuits [63] and acoustic systems [64], has not yet been realized in optics. This state emerges due to even stronger NNN coupling, introducing additional connections between edge and bulk cells.

We find that the S-type-III corner state almost mixes with the A-type-II corner state, as manifested in the eigenstates between type-I and A-type-III states in the simulated results (Fig. 2b). The tight-binding model fails to explain this phenomenon due to the structural complexity of the lattice. To explore this mixing further, we introduce symmetry considerations and argue that it arises from the parity engineering of edge-edge couplings in our alternating trivial-topological Kagome lattice. The entire lattice can be classified into three types of units: single isolated sites, dimers located at the domain walls, and trimers within the bulk (Fig. 2a). Typically, type-II corner state are spectrally separated from the edge states due to edge-edge coupling, which primarily occurs via dimer-dimer interactions. In contrast, type-III corner states rely on coupling between dimers and trimers. In this configuration, six dimers located at the edges and nine trimers between them participate in the coupling interactions, as illustrated along each hexagon encircled by a dashed circle in Fig. 2c. Moreover, coupling between dimers and trimers in the lattice should be taken into account to provide an effective framework for understanding the symmetry of corner states with respect to the electric field charge distribution, in which the trimers act as electric monopoles and dipoles in different scenarios. For the antisymmetric mode, the $C_3$ symmetry allows the six type-III dimers along the dashed circle trajectory to exhibit an alternating distribution of positive and negative charges. The coupling between these boundaries is mediated by trimers through parity engineering: an odd number (one) or even number (two) of trimers can connect two edge dimers. We label the six dimers as 1-6, then assign each dimer either a positive or negative electric charge distribution. Arranging them in an antisymmetric pattern from 1 to 6, we find that the trimer can function as either an electric monopole or an electric dipole to connect the coupling between dimers (top panel in Fig. 2c).

However, a symmetric charge distribution is not allowed. We can assign all six dimers with charge distributions in two scenarios: Scenario 1(bottom left panel in Fig. 2c) or Scenario 2, (bottom right panel in Fig. 2c). In both cases, each of the three topological lattices (shaded area) can exhibit a symmetric profile. When a trimer functions as an electric monopole, it must have an opposite charge distribution to couple with another trimer. Conversely, when a trimer acts as an electric dipole, it must couple with another trimer that has an opposing charge distribution. Following this rule, we find that neither a monopole nor a dipole charge distribution is allowed in Scenario 1, as the charge distributions in the trimers between Dimer 2 and Dimer 3 (also between Dimers 4 and 5, 6 and 1) are not symmetry-broken. Similarly, in Scenario 2, a contradictory charge arrangement arises between Dimer 6 and Dimer 1. These forbidden couplings are indicated by the cross marks in Fig. 2c. As a result of this parity constraint, the A-type-III corner state successfully stands out and remains spectrally isolated from the edge states, whereas the S-type-III corner state fails due to parity incompatibility. In contrast, the S-type-III corner state is more likely to remain stable against edge-state mixing within a single topological Kagome lattice (*SI Appendix*, section 2). This is because only a single trimer functioning as a monopole is required in such a configuration. Notably, while our design involves three topological regions, the S-type-III state becomes unstable due to parity engineering. In contrast, the A-type-III state remains exceptionally stable under the same conditions. These results highlight the ability of parity engineering to precisely control corner states in photonic systems. Importantly, such findings have not been reported in any other platforms to date.

The simulated electric field distribution profiles of the corner states, corresponding to the eigenstate calculations in Fig. 2b, are plotted in Fig. 2d. Clearly, both type-I and type-III corner states are observed, with the highest field intensities localized at the center pillar and the second-nearest dimers, respectively. Notably, the antisymmetric field distribution of the $E_z$ component with respect to the corner bisector confirms the presence of A-type-III corner state-- an elusive mode that, until now, has remained unobserved in photonic systems.

**Experimental Characterization**

To characterize the lasing action, we used a 355 nm nanosecond laser pulse to pump the center of the device and collected the emitted photons. The pump-fluence dependence of the PL intensity spectra was investigated, with the measured results shown in Fig. 3a. As the pump intensity increases, a clear transition from spontaneous emission to stimulated emission is observed, marked by the sequential emergence of two sharp lasing peaks. Specifically, a distinct lasing peak at 524.01 nm appears first when the pump fluence reaches 74 µJ/cm². With further increase in pump fluence, three additional lasing peaks appear. A close-up inspection at a pump fluence of 102 µJ/cm², shown in Fig. 3b, reveals four peaks at 524.01 nm, 526.32 nm, 528.05 nm and 530.45 nm, respectively. The A-type-II and S-type-III corner states contribute little to the lasing, as indicated by the small bump between the type-I and type-III corner lasing peaks (Fig. 3b). These experimental results align with our 3D first-principles FEM simulations in Fig. 2b, which predict the type-I corner state at 576.16 THz (~ 520.69 nm) and the A-type-III corner state at 566.77 THz (~ 529.32 nm). As noted earlier, the fabricated sample exhibits a narrower bandgap compared to the simulation.

As shown in Fig. 3c-f, the nonlinear increase in emission intensity confirms the onset of lasing, with the lasing thresholds of the four modes identified at the "kink" points at approximately 74.0 µJ/cm², 77.2 µJ/cm², 74.2 µJ/cm² and 81.4 µJ/cm², respectively. Furthermore, the evolution of linewidth as a function of pump fluence reveals a significant narrowing from ~32 nm to within 1 nm at the thresholds, further verifying the lasing behaviour of the designed nanocavity. The experimental results show excellent agreement with the numerical predictions in Fig. 2b, confirming that the first emerging lasing peak corresponds to the type-I corner state, while the fourth peak originates from the A-type-III corner state, which has so far evaded observation in photonic systems. Importantly, detecting the A-type-III corner state poses a significant challenge not only in photonics but also in non-photonic platforms such as electrical circuits [61]. Nevertheless, our design effectively engineers edge-edge coupling and its parity, enabling experimental detection of this elusive corner state through lasing action.

Notably, the measured results demonstrate the robustness of topologically protection within the NIL scheme, as the corner states remain robust against the fabrication imperfections clearly shown in Fig. 1c. Here we present results from a sample featuring pre-designed defects on the nanoimprint template. Another sample fabricated using a different template was also characterized (*SI Appendix*, section 6). We note that A-type-II and S-type-III corner states remain mixed; however, these two states can be separated in a structure containing only a single topological Kagome lattice, whereas A-type-III corner state remains undetectable (*SI Appendix*, section 2). Moreover, type-I and A-type-III corner states are consistently observed across different cases, with only minor frequency shifts in the spectra. This highlights the importance of parity engineering in topological photonics for realizing specific topological states. These results further demonstrate how topological design principles enhance the applicability of NIL fabricating strategies.

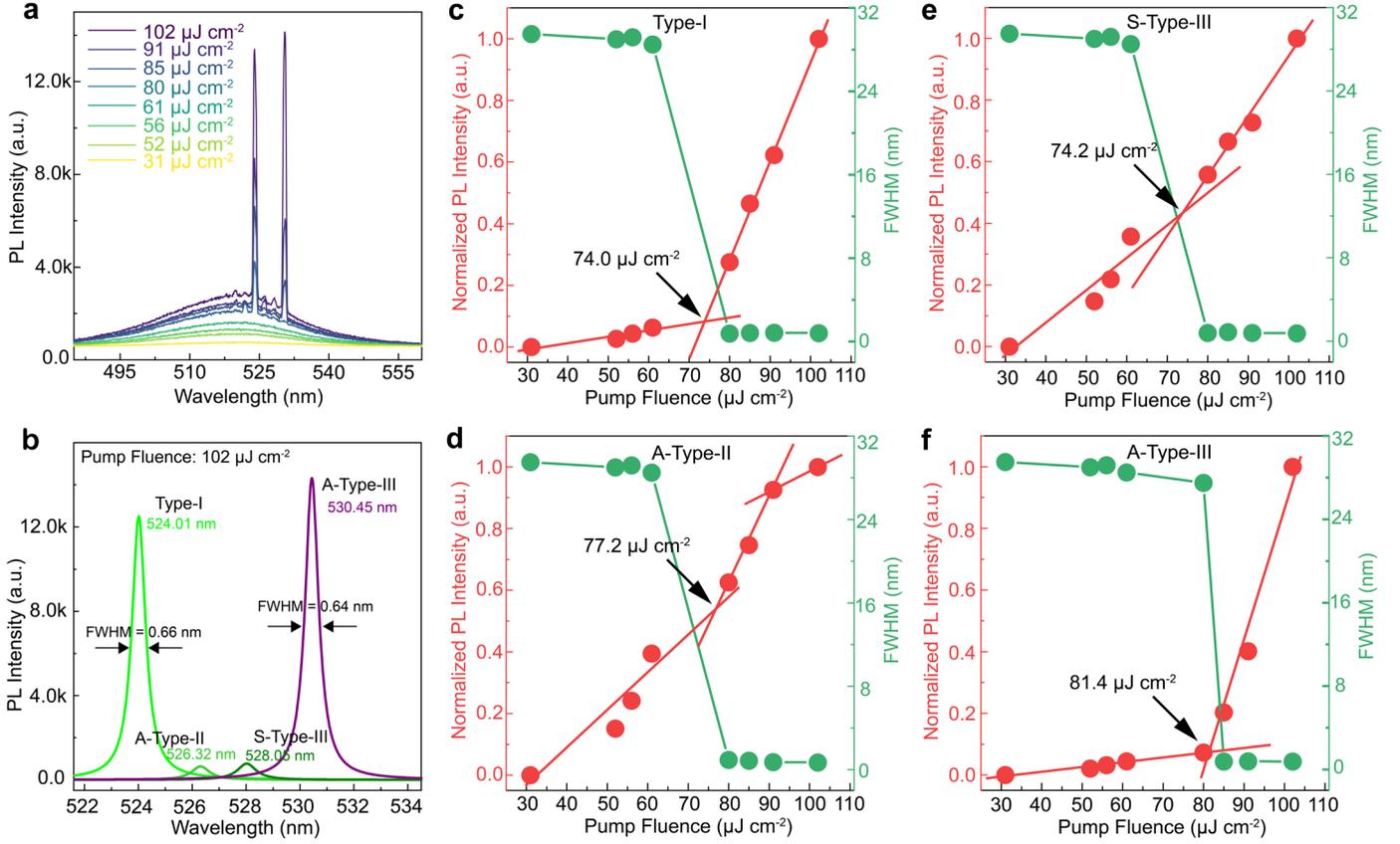

**FIG. 3. Characterization of the lasing behaviour of multiple corner states. (a)** Measured PL spectra under different pump fluences. **(b)** Close-up PL spectrum of the resonance peak at corner states under pump fluence of 102 µJ/cm². **(c-f)** Light-in light-out curve and the evolution of linewidth of the corner states as a function of pump fluence, substantial narrowing of linewidth at a threshold of 74.0 µJ/cm², 77.2 µJ/cm², 74.2 µJ/cm² and 81.4 µJ/cm² for different types of HOTCS, respectively.

## Summary

In summary, we have successfully demonstrated a one-step, handmade, nanoimprinted topological laser operating with multiple HOTCS. This low-threshold laser emits in the green region of the visible spectrum at room temperature. By implementing a modified Kagome lattice of coupled perovskite nanopillars, our work represents the first realization of a type-III corner state in photonic systems. We characterized the topological corner states through lasing action detection, theoretical tight-binding calculations, and full-wave first-principles simulations. Most importantly, we achieved modulation of symmetric- (S-type) and antisymmetric- (A-type) type-III corner states via parity engineering. The topological laser exhibited remarkable robustness against fabrication imperfections in the nanoimprinting process. Our work not only demonstrates the significant potential of nanoimprinting technology for the large-scale production of topological photonic devices using soft materials, but also extends the scope of higher-order topology to visible photonic systems. Furthermore, it provides valuable insights into how functional robustness can be introduced into conventional NIL fabrication strategies through the integration of topological design.

## Materials and Methods

**Measurement of Steady-state Photoluminescence (PL), Absorption, Photoluminescence Quantum Yield (PLQY)** For steady-state PL/absorption measurements, a home-built optical setup was employed. Light sources included a halogen tungsten lamp for absorption measurements and a He-Cd laser for PL measurements. Light signals were resolved and recorded by a monochromator equipped with a photomultiplier, using a standard lock-in detection technique to improve the signal-to-noise ratio. The monochromator's diffraction grating was set to 600 grooves per mm, and the measurement step was set at 0.2 nm. The PLQY of the purified samples dispersed in hexane was measured using the FS5 Spectrofluorometer from Edinburgh Instruments Ltd, which includes an integrating sphere optical system and a monochromator.

**Preparation of the Original Master Structure** The original silicon master was purchased from CONSCIENCE, Sweden. It is constituted of arrays of circular holes in a Kagome lattice shape with 190 nm depth (designed 20 nm deeper than the simulation result to leave space for the air bubble residuals) and 278 nm periodicity ($\sqrt{3}a$, where $a = 160$ nm is the side length of the Kagome hexagon). For the single circle in the center, the diameter is 90 nm. Each nanopillar is shrank/expanded towards/backwards the central point of the hexagon by a distance of $0.15a$. The surface of the silicon master was treated for anti-adhesion using gas phase deposition of silane chemistry. Specifically, a drop of perfluorooctyltrichlorosilane was added near the silicon master in a Petri dish within a nitrogen chamber, and then the dish was sealed. The entire system was left for 6 hours to complete the deposition process. Afterwards, the silicon master was rinsed with acetone to remove any unreacted silane. This anti-adhesion treatment will facilitate the demolding of subsequent intermediate molds.

**Preparation of the Working Masters (Negative Replica of the Silicon Master)** The intermediate mold is a negative replica of the silicon master, created using nanoimprint lithography. During the preparation of the glass substrate, OrmoPrime20 is used to enhance the adhesion between OrmoStamp and the glass substrate. Specifically, for substrate treatment, the slides are first cleaned with ozone plasma to ensure they are free of impurities and moisture, achieving optimal adhesion of OrmoPrime20. The slides are then baked at 195°C for 30 minutes and cooled to room temperature immediately before spin-coating OrmoPrime20 to obtain a uniform layer. Finally, the slides are baked on a hot plate at 85°C for 2–3 minutes to remove residual solvents from the film and cure the OrmoPrime20.

In low light conditions, a drop of the polymer photoresist OrmoStamp is applied directly to the surface of the silicon master. The prepared glass substrate is then immediately pressed gently onto the polymer photoresist, ensuring no air bubbles remain. An ultraviolet curing lamp is used to cross-link and cure the resist for 3 minutes. Finally, the silicon master and glass substrate are transferred to a hot plate, where the temperature is raised to 180°C and maintained for 2–3 minutes. The thermal expansion effect causes the substrate and silicon master to separate automatically. Due to the enhanced adhesion provided by OrmoPrime20, the OrmoStamp layer adheres to the glass slide after demolding.

**Preparation of PDMS Imprinting Mold** A PDMS mold was prepared using a combination of GELEST® hPDMS and Sylgard 184 PDMS to better preserve nanostructures during the imprinting process. Specifically, GELEST® hPDMS Part A and Part B were mixed in a 1:1 ratio. The hPDMS mixture was then dispensed onto the working master using a pipette, and an air gun was used to carefully remove any bubbles to ensure the hPDMS evenly covered the entire structure. Next, the working master was transferred to a hot plate and heated at 80°C for 4 hours to cure the hPDMS.

After curing the hPDMS layer, Sylgard 184 PDMS was prepared by mixing its monomer and crosslinker in a 10:1 weight ratio. The Sylgard 184 PDMS mixture was poured over the already cured hPDMS. The combined layers were then cured at 85°C for 4 hours. Finally, the PDMS imprinting mold was manually demolded from the master.

**Preparation of DBR Substrate** To enhance the substrate reflectivity, a DBR mirror consisting of alternating layers of $SiO_2$ and $TiO_2$ was selected, with a design centered around a wavelength of ~530 nm. The DBR was ultrasonically cleaned in a mixed solution of isopropanol and acetone, followed by hydrophilic treatment using a plasma cleaner.

**Self-Assembly of Topological Nanopatterns** Initially, a centrifuge is used to remove excess surfactants from the NCs, followed by the use of hexane to disperse the NCs. Subsequently, 1 μL of the NCs solution is deposited onto the pre-treated DBR substrate. The PDMS mold is then promptly positioned over the solution, and allowed to dry in a nitrogen cabinet under dark conditions for 120 minutes. Finally, the PDMS mold is carefully removed to reveal the imprinted pattern of NCs.

**Numerical simulation** All simulations are performed using the optics module (Electromagnetic Waves, Frequency Domain) of COMSOL Multiphysics, which is a commercial software based on the Finite Element Method. In the 3D simulation, we used $n = 2.0$ as the refractive index of the nanopillars. Each nanopillar has a diameter of 90 nm. The nanopillar is shrank/expanded towards/backwards the central point of the hexagon by a distance of $0.15a$ in each unit cell. The height of the nanopillars is 170 nm. Perfect electric conductor (PEC) condition is applied to the bottom surface to emulate the reflector.

## Acknowledgements

This work was supported by NRF-CRP23-2019-0007 and Tier 1 (MOE-RG139/22). H. D. Sun acknowledges the support from CPG2024-00006, SRG2023-00025, and the Science and Technology Development Fund (FDCT), Macao SAR (File no. 0122/2023/RIA2). X. C. Yuan acknowledges the support from Guangdong Major Project of Basic and Applied Basic Research (2020B0301030009). Q. Zhang acknowledges the support from National Natural Science Foundation of China (12104318).

## Author contributions

Q. Z., R. D., B. L. Z. and H. D. S. conceived the idea. Q. Z. and R. D. designed and discussed the experiments. R. D. fabricated the samples and performed the optical characterization experiments. L. W. and X. Y. contributed to the fabrication of perovskite NCs. Q. Z. did the theoretical analysis, with R. D. and A. Y.'s input. Q. Z., R. D., B. L. Z. and H. D. S. analysed the data. All authors participated in the discussions. Q. Z., R. D., B. L. Z. and H. D. S. drafted the manuscript. H. D. S., B. L. Z. and R. D. supervised the research.

## Competing interests

The authors declare no competing interests.